\pdfoutput=1 
\documentclass[10pt,aps,prb,amsmath,amssymb,twocolumn,%
				letterpaper,showpacs,footinbib,%
				balancelastpage,raggedbottom,
				citeautoscript,reprint,longbibliography,%
				floatfix]{revtex4-1}
\usepackage[usenames,dvipsnames]{xcolor}
\usepackage[bookmarks=false,colorlinks]{hyperref}
\hypersetup{
    linkcolor=RubineRed,        
    citecolor=ForestGreen,     
    filecolor=Mulberry,      	
    urlcolor=RoyalBlue,           
}
\usepackage{graphicx}
\usepackage[protrusion=true,expansion=true,final]{microtype}
\usepackage{xfrac}
\usepackage[scaled=0.98]{helvet} 
\usepackage{xspace}

\usepackage{ccaption}
\usepackage{ragged2e}
  \captionnamefont{\small \bfseries \sffamily}
  \captiontitlefont{\small \sffamily }
  \captiondelim{~\textbar~}
  \captionstyle{\justifying}



\newcommand{\scro}{(Sr,Ca)Ru$_2$O$_6$\xspace}

\begin{document}
\title{Designing a robustly metallic noncenstrosymmetric ruthenate oxide with large thermopower anisotropy 
}
\author{Danilo\ Puggioni}
	\affiliation{Department of Materials Science \& Engineering,\!
	Drexel University,\! Philadelphia,\! PA 19104,\! USA}%
\author{James M.\ Rondinelli}
  \email{jrondinelli@coe.drexel.edu}
	\affiliation{Department of Materials Science \& Engineering,\!
	Drexel University,\! Philadelphia,\! PA 19104,\! USA}%
%
%

%
\begin{abstract}
The existence of approximately 30 noncentrosymmetric metals (NCSM) 
suggests a contraindication between crystal structures 
without inversion symmetry and metallic behavior. 
Those containing oxygen are especially scarce. 
Here we propose and demonstrate a design framework to remedy this  
property disparity and accelerate NCSM-oxide discovery: 
The primary ingredient relies on 
the removal of inversion symmetry through 
displacements of atoms whose electronic degrees of freedom 
are decoupled from the states at the Fermi level.
Density functional theory calculations validate this crystal--chemistry  
strategy, and we
predict a new polar ruthenate 
exhibiting robust metallicity. 
We demonstrate that the electronic structure
is unaffected by the inclusion of spin-orbit interactions (SOI), and 
that cation ordered SrCaRu$_2$O$_6$ exhibits a large thermopower 
anisotropy ($\left|\Delta\mathcal{S}_\perp\right|\sim6.3~\mu \textrm{V K}^{-1}$ at 300~K) derived from 
its polar structure. 
Our findings provide chemical and structural selection guidelines 
to aid in the search of new NCS metals with enhanced thermopower anisotropy.
\end{abstract}
%
\maketitle
\sloppy

\begin{figure}[t]
\centering
	\includegraphics[width=0.49\textwidth,clip]{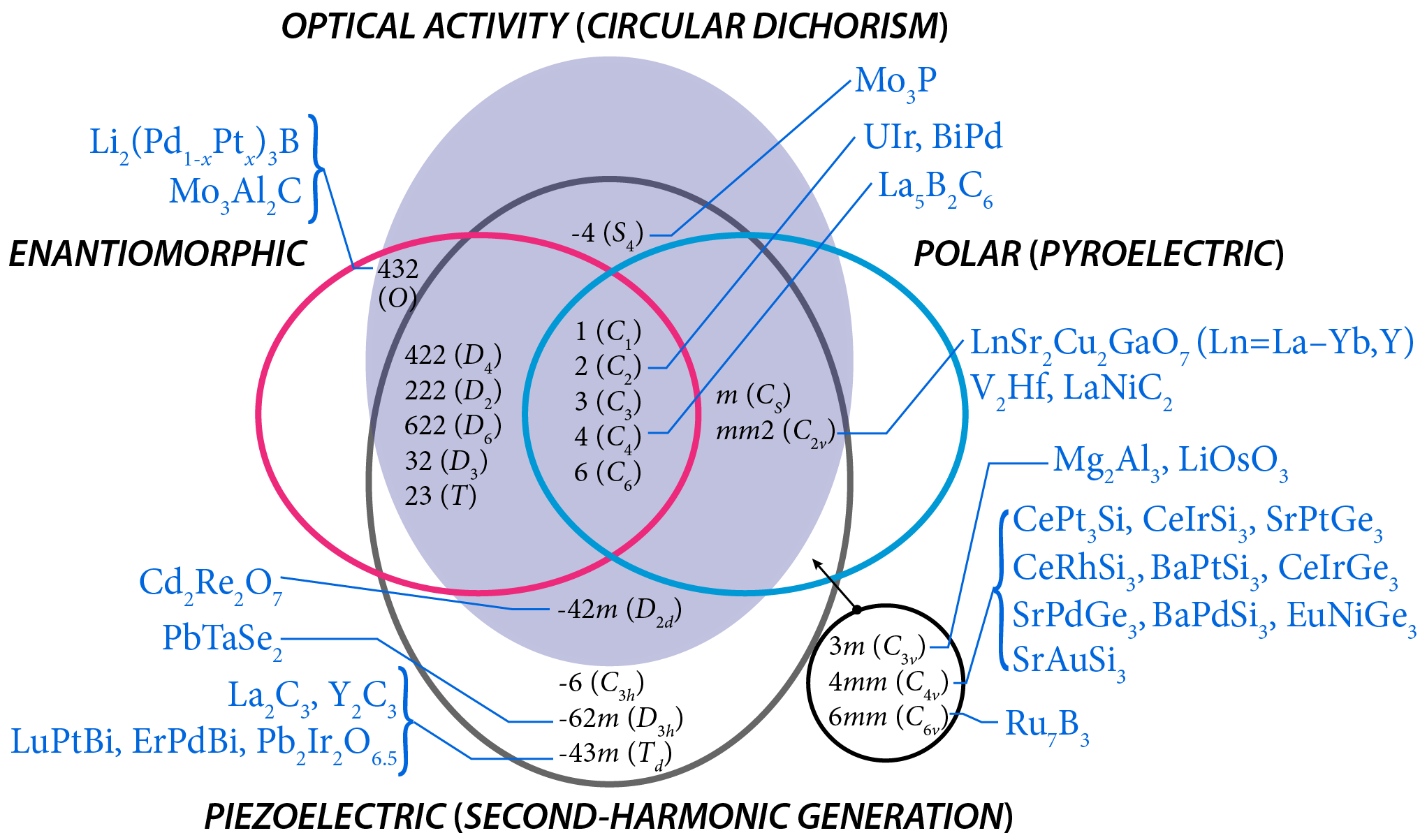}\vspace{-0.7\baselineskip}
	\caption{\textbf{Classification of noncentrosymmetric metals.}
	Known compounds differentiated by crystal class 
	following the NCS classification scheme introduced in  
	Ref.~\onlinecite{Halasyamani/Poeppelmeier:1998},   
	include  4$d$ and 5$d$ transition metal intermetallics, carbides or silicides.
	Oxide compounds are scarce.
	}
	\label{fig:ncs_metals}
\end{figure}

The metallic features in materials, which  provide low-resistance channels for 
electrical conduction, lead to effective screening of local electric dipole moments.\cite{Lines/Glass:1977} 
Itinerant electrons disfavor both their formation and 
cooperative ordering.\cite{Zhong/Vanderbilt/Rabe:1994} 
In spite of the incompatibility between acentricity and metallicity, metallic materials which break 
the spatial parity operation mapping $(x,y,z)\rightarrow(\bar{x},\bar{y},\bar{z})$ 
were originally discussed in the 1960s by Matthias 
and then later more rigorously by Anderson and Blount.\cite{Anderson/Blount:1965}
The first experimental identification\cite{Lawson:1972} of a candidate polar metal, 
the binary intermetallic V$_2$Hf, was made a decade later.
Since then, more noncentrosymmetric metals (NCSM) 
have been identified (\autoref{fig:ncs_metals}) and found to 
exhibit unconventional optical responses,\cite{Mineev/Yoshioka:2010,Edelstein:2011} 
magnetoelectricity\cite{Edelstein:1995,Edelstein:2005} 
and superconductivity.\cite{Edelstein:1996,Bauer/Feuerbacher_etal:2007,Bauer/Rogl_etal:2010}
Yet, they remain challenging to discover. 
The principal NCSM material classes are binary and ternary intermetallics and silicides.
Intriguingly, few NCSM are {oxides} with the notable exceptions\cite{Poeppelmeier_etal:1991,Vaughley/Poepelmeier:1991,Sergienko/Mandrus:2004,Boothroyd:2013} 
of a layered cuprate (Y$_{1-x}$Ca$_x$Sr$_2$GaCu$_2$O$_{7\pm y}$),  
a geometrically frustrated pyrochlore (Cd$_2$Re$_2$O$_7$), 
and recently LiOsO$_3$ with a structural transition similar to ferroelectric LiNbO$_3$. 
(Degenerately doped dielectrics like BaTiO$_{3-\delta}$ \cite{kolodiazhnyi:2010} and ZnO:Al,\cite{dasamitk:2012}
although exhibiting metallic resistivity, are not intrinsically metallic {and} 
NCS---their conductivity is high because of deviations from ideal stoichiometry.)
\autoref{fig:ncs_metals} suggests  
NCSM tend to require cations with large atomic masses, but this 
is not always the case, \emph{viz.}, Mg$_2$Al$_3$. 
The absence of reliable crystal-chemistry guidelines and the limited 
understanding  of the microscopic origin of inversion symmetry lifting 
displacements in metals poses a serious challenge for their discovery: 
{How does one explain, let alone design, cooperative acentric atomic 
displacements in crystalline oxide conductors?}  
In this work, we present a microscopic understanding for noncentrosymmetric metals, leading to a design principle which we use in the theoretical prediction of a 
polar-NCSM transition metal (M) oxide.
We formulate a two-step materials selection process based on a 
weak electron--lattice coupling principle that  
considers the symmetry and atomic species participating in the 
inversion lifting lattice modes, and the susceptibility of the cations' 
electronic configurations to a metallic state.
We show that two $k\neq0$ lattice modes describing rotations 
of MO$_6$ octahedra induce polar A-cation displacements in the layered and 
cation ordered (Sr,Ca)Ru$_2$O$_6$ perovskite, for which 
experimental 
solid solutions\cite{Cao_et_al:1997} are  metallic.
Because the microscopic mechanism  is 
independent of the details of the states at the Fermi level, 
 unlike conventional dipolar interactions  which are essential to stabilizing acentric 
displacements of  second-order Jahn-Teller cations in dielectrics,\cite{Wang/Tsymabl:2012} the 
design strategy is readily transferable to other transition metal chemistries.
We show that spin-orbit interactions do not alter the electronic ground state, and 
that the unusual coexistence of a polar axis and metallicity produces 
a novel correlated material possessing a highly anisotropic 
thermopower response ($\left|\Delta\mathcal{S}_\perp\right|\sim6.3~\mu  \textrm{V K}^{-1}$ at 300~K) 
with nearly isotropic electrical conductivity.\\

\begingroup
\squeezetable
\begin{table*}[t]
\begin{ruledtabular}
\centering
\caption{\label{tab:ansatz}\textbf{\sffamily Routes to lift inversion symmetry in metallic compounds.}
The scheme is formulated by classifying the symmetry 
requirements of the  lattice mode instabilities of a centrosymmetric 
(reference) phase which would 
drive a thermodynamic transition to a noncentrosymmetric crystal structure. 
Although compositional ordering may be an obvious route to lift inversion 
symmetry, we recognize that experimentally in practice it may be challenging to achieve 
it, especially in oxide materials owing to bond-coordination requirements; therefore, 
although the route is simple, it is by no means straightforward.
The design methodology demonstrated here implements method (3) (in boldface).
Note that all materials are expected to show some interesting physical properties enabled 
by an acentric crystal structure with itinerant electrons, independent of the mechanism 
leading to the mutual coexistence of the prerequisite atomic and electronic structure.} 
\begin{tabular}{lp{1.35in}lp{3.95in}}%
\multicolumn{2}{l}{\sffamily \bfseries Inversion-lifting Method} & \sffamily \bfseries Mode Requirements & \sffamily \bfseries Description\\
\hline
\sffamily (1)& \sffamily Compositional Order  
	&  \sffamily None 
	&  \sffamily Achieved by decorating one or more interleaved lattices with multiple cations, 
	{e.g.}, as realized in half-Heulser alloys or tri-color, 
	$\cdot$/ABC/ABC/$\cdot$,  superlattices.\\[0.4em]
\sffamily (2)& \sffamily Packing of Acentric Polyhedra\raggedright\let\\\tabularnewline
	&\sffamily  1, $k=0 \textrm{ or } k\neq0$ mode
	& \sffamily Realized by the alignment of acentric metal-oxygen polyhedral units arising from 
	either out-of-center metal distortions or because they 
	intrinsically lack inversion (MO$_4$ tetrahedron), {and} 
	contribute few (if any) states to or near the Fermi level. \\[0.4em]
\sffamily \textbf{(3)}& \sffamily \textbf{Geometric-induced  Displacements}\raggedright\let\\\tabularnewline 
	& \sffamily $\ge~\!$2,  $k\neq0$ coupled modes 
	&  \sffamily Obtained by anharmonic coupling of two or more centric lattice modes, which cooperatively remove inversion and produce cation displacements that do not gap the electronic structure. 	Accessible in some, effectively, two-dimensional compounds, {e.g.},  naturally layered structures or 3D  systems with bi-color ordering.\\
\end{tabular}
\end{ruledtabular}
\end{table*}
\endgroup

\noindent{\bfseries \sffamily Results}\\\vspace{-1\baselineskip}

\noindent\textbf{Design of a metallic oxide without inversion.} %
Our design strategy originates from Anderson's work\cite{Anderson/Blount:1965} on ``ferroelectric metals,''
where he writes that, ``while free electrons screen out the electric field [in materials] completely, 
they do not interact very strongly with the transverse optical phonons and the 
Lorentz local fields [that] lead to ferroelectricity, since umklapp processes are 
forbidden as $k\rightarrow0$.''
We recast this observation into an {operational principle}  that states:  
{The existence of any NCSM relies on weak coupling 
between the electrons at the Fermi level, and the (soft) phonon(s) 
responsible for removing inversion symmetry.}
%
An essential, implicit, materials constraint is that the low-energy electronic structure 
derives from an electron count giving partial band occupation, which may be obtained by 
judicious selection of the cation chemistries.
We identify three displacive routes fulfilling the weak-coupling hypothesis 
in solid-state systems,
specified by the symmetry behavior of the lattice phonons that would 
remove inversion symmetry in a centrosymmetric metal and yield a 
NCS crystal structure (\autoref{tab:ansatz}). 
It should be noted that an order--disorder mechanism could also be 
operative; the classification of such symmetry breaking is consistent with 
our scheme; the caveat being that the soft phonon is replaced by an 
order parameter that  describes atomic site occupancy. 
In either case the same set of irreducible representations may be used.
The third approach, which we focus on here, resembles a condition which supports a novel 
``improper'' mechanism\cite{Benedek/Fennie:2011,Bousquet/Ghosez_et_al:2008} for ferroelectricity 
that has been exploited to design artificial polar oxides with sizeable electric polarizations 
by heterostructuring non-polar dielectrics. 
Note that throughout we choose to refer to these materials as {NCS metals}  
rather than ``ferrroelectric-metals'' as articulated by Anderson and recently others to eliminate misconceptions 
about the presence of switchable spontaneous electric polarizations in metals, and 
more accurately describe the crystallographic--electronic function of the materials.
In this way, one can generally and completely classify 
all metals without inversion symmetry and not solely those of the polar subset of 
NCS structures (\autoref{fig:ncs_metals}).
Within this framework, the design of a NCSM oxide using 
geometric-induced displacements requires 
an oxide  class exhibiting coupled zone-boundary phonons that 
lift inversion symmetry.
Here we choose {orthorhombic} AMO$_3$ perovskites  
with corner-connected MO$_6$ octahedra that are rotated about each Cartesian axis; 
the ``tilt'' pattern is obtained and described by two phonons with $k\!=\!(\sfrac{1}{2},\sfrac{1}{2},0)$ and $(\sfrac{1}{2},\sfrac{1}{2},\sfrac{1}{2})$ wavevectors relative to the cubic aristotype.\cite{Howard/Stokes:1998}
The $a^+b^-c^-$ MO$_6$ rotation pattern in the presence of {layered} A-cation order along an [001]-direction, e.g., accessible through  synthetic 
bulk chemistry routes\cite{Graham/Woodward:2010,Dachraoui/Greenblatt:2011} or 
heteroepitaxial thin film growth methods\cite{Zubko/Triscone_et_al:2011}, 
is sufficient to produce a polar structure.\cite{Rondinelli/Fennie:2012,Mulder/Rondinelli/Fennie:2013}
The chemical species are selected given the constraints that the rotations of 
octahedra are necessary%
---determined by the A and M cation size mismatch in perovskites---and 
that the electronic configuration of the M cation results in a nonzero  
density-of-states at the Fermi level ($E_\mathrm{F}$). 
These requirements should be satisfied for ordered (Sr,Ca)Ru$_2$O$_6$, since 
both bulk SrRuO$_3$ and CaRuO$3$ are experimentally found in the 
$a^+b^-c^-$ Glazer tilt pattern and also have partially occupied 
low-energy Ru 4$d$ $t_{2g}$ orbitals.\\

\begin{figure}
	\centering
  \includegraphics[width=0.97\columnwidth,clip]{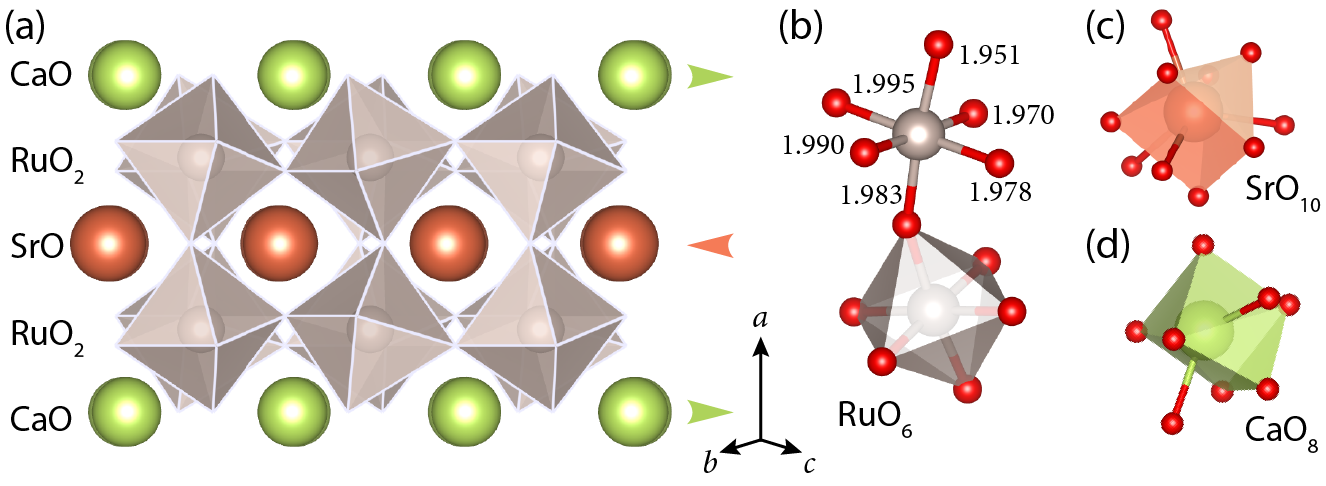}\vspace{-1\baselineskip}
  \caption{\textbf{Ground state structure of the polar noncentrosymmetric metal (Sr,Ca)Ru$_2$O$_6$.} 
   Crystal structure of cation ordered (Sr,Ca)Ru$_2$O$_6$ 
   with the $a^+b^-b^-$ RuO$_6$  tilt pattern (a).
  The off-center  Ru$^{4+} $distortions (bond lengths
  in \AA) produce local dipoles which cooperatively 
  order along the $c$-direction (b), removing the mirror plane 
  perpendicular to the A-site cation order. 
  The local coordination environment for (c) Sr (tetra-capped trigonal prism) 
  and (d) Ca (bi-capped trigonal prism) reveal that large A-site 
  distortions from the ideal 12-fold coordination occur in opposite directions (arrows), 
  along the polar $c$-axis.} 
  \label{fig:structure}
\end{figure}

\noindent\textbf{Ground-state structure.}
\autoref{fig:structure} depicts the polar $Pmc2_1$ ground state crystal structure of 
(Sr,Ca)Ru$_2$O$_6$ obtained from density functional theory (DFT) calculations. 
(See Supplementary Table 1 for the crystallographic data.)
The extended RuO$_6$ structure is highly rotated, 
adopting the targeted orthorhombic  tilt pattern with 
%
rotations of adjacent octahedra out-of-phase about the 
[011]-direction and in-phase about the layering [100]-direction [\autoref{fig:structure}(a)].
The Ru--O--Ru bond angles are 157.1$^\circ$ and 145.6$^\circ$ (perpendicular to the 
CaO/SrO layers) and 150.4$^\circ$ (within the layers).
The stabilization of this rotation pattern is understood on 
the grounds that the $\pi^*$ band, which is nearly full because of 
the $t_{2g}^4$ configuration, is shifted to lower energy by the 
electronegative Ru$^{4+}$  cation.\cite{Woodward:1997b}

(Sr,Ca)Ru$_2$O$_6$ exhibits small off-center Ru distortions 
[\autoref{fig:structure}(b)] and large A-site cation displacements  [\autoref{fig:structure}(c,d)]. 
Acentric B-site displacements are common to 
TMs with $d^0$-electronic configurations in 
octahedral coordinations; however, for $d^n, n\geq1$, 
such polar displacements are largely disfavored, {i.e.}, 
the energetic gain due to $d\pi-p\pi$ metal--oxygen bonding 
decreases upon filling the $t_{2g}$ orbital.\cite{Kunz/Brown:1995} 
Nonetheless, we find they occur for  $d^4$ Ru$^{4+}$, displacing the metal 
center nearly towards an edge ($\sim$0.03~\AA), staying in the 
(110) plane (see Supplementary Table 2).  
This results in one short-, 
two long-, and three medium Ru--O bonds [\autoref{fig:structure}(b)], which 
are close to the average value found experimentally in the Sr$_{0.5}$Ca$_{0.5}$Ru$_2$O$_3$ 
solid solution.\cite{Kobayashi_etal:1994}  
The Ca and Sr atoms make large polar displacements (0.29 and 0.21~\AA, respectively) 
along the [001] and [00$\bar{1}$] directions (see Supplementary Table 3), 
resulting  in distorted polyhedra  [\autoref{fig:structure}(c)].
%
We explore these displacements and their effect on the electronic structure in more detail below.\\

\begin{figure}
	\centering
  \includegraphics[width=0.99\columnwidth,clip]{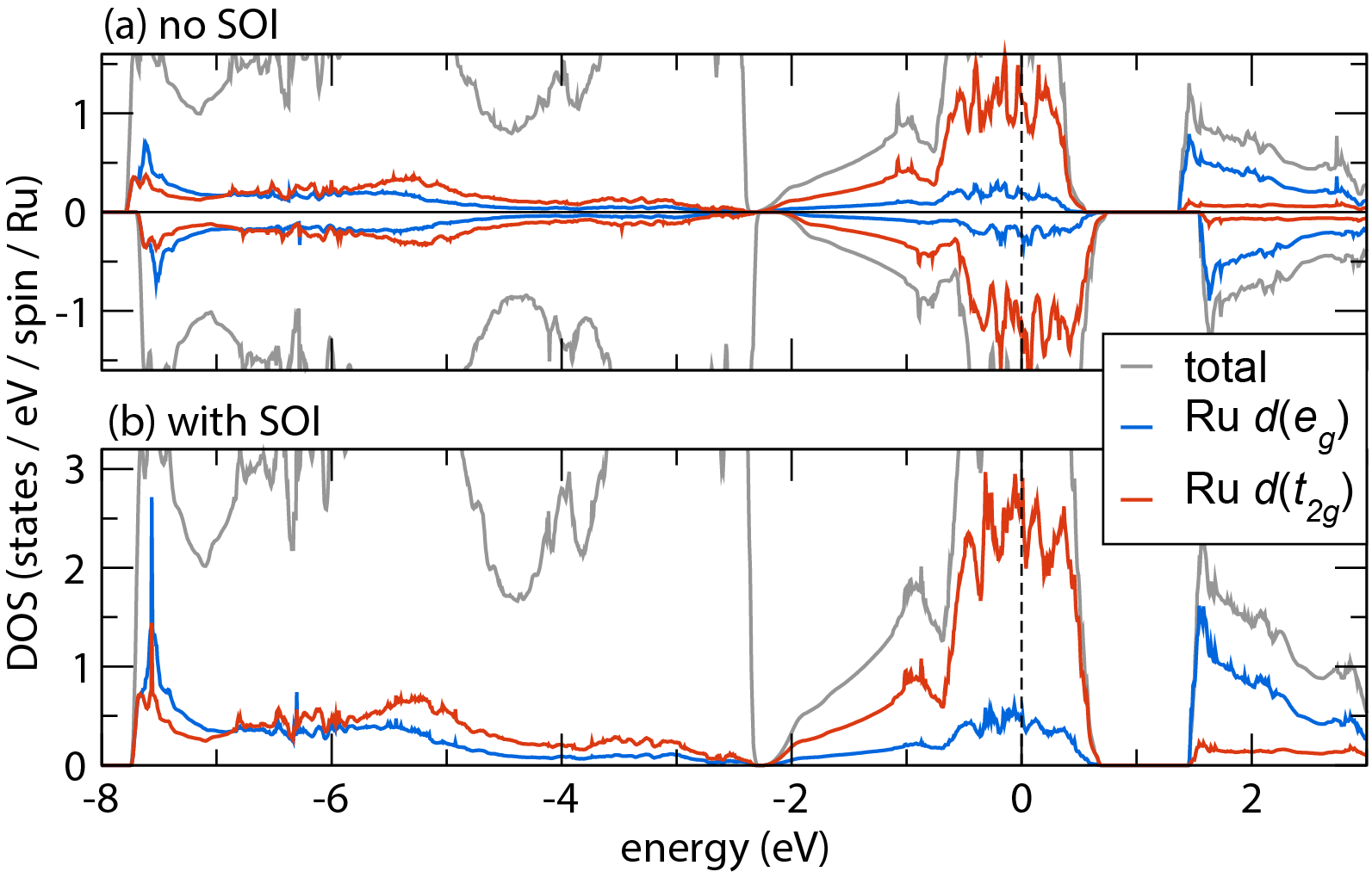}\vspace{-1\baselineskip}
  \caption{\textbf{Electronic structure.} The total (black line) and Ru 4$d$ orbital- and 
  spin-resolved densities of states (DOS) orthorhombic (Sr,Ca)Ru$_2$O$_6$ 
  calculated (a) without spin-orbit interactions (SOI) and (b) with 
  SOI. The Fermi level is at 0 eV (dashed line).}
  \label{fig:dos}
\end{figure}

\noindent\textbf{Electronic properties.}
\autoref{fig:dos} shows the total and partial densities-of-states 
(DOS) for the polar-NCSM (Sr,Ca)Ru$_2$O$_6$ without (panel a) 
and with (panel b) spin-orbit interactions (SOI).
%
The valence band is composed largely of O 2$p$ states hybridized with Ru 4$d$ 
states, with the oxygen states predominately found in 
regions below the $E_\mathrm{F}$.
Consistent with  low-spin 
Ru$^{4+}$, the large peaks in the DOS near  $E_\mathrm{F}$  are 
caused by the fairly flat Ru $t_{2g}$ bands, whereas the
strongly hybridized $e_g$ orbitals form broader bands in the conduction band 
beginning near 1.5~eV.
We find that without SOI, weakly dispersive $t_{2g}$
bands lead to a large number of states at the 
Fermi level, {i.e.}, $N(E_\mathrm{F})\sim2.2$~eV/spin/Ru atom\ indicating 
robust metallic behavior [\autoref{fig:dos}(a)], despite 
a $\sim$0.75~eV band gap present in both spin 
channels far above $E_\mathrm{F}$ (0.5~eV).
It is also ferromagnetic (0.914~$\mu_\mathrm{B}$/f.u.)\ with 
$\sim$0.35~$\mu_\mathrm{B}$/Ru atom; the remaining magnetization is  
distributed among the oxygen ligands due to the 
itinerant spin-polarized electrons.
Using a simple modified Weiss formula,\cite{Lampis:2004} we calculate a Curie temperature
$T_c=47$~K (27~K applying Anderson's rescaling\cite{Anderson:1959}), 
in good agreement with experimental data\cite{Cao_et_al:1997}  on 
solid solution Sr$_{0.5}$Ca$_{0.5}$RuO$_{3}$  ($T_\mathrm{c}=57$~K). 

Since spin-orbit interactions play a significant role in determining the orbital structure in 
3$d$ transition-metal oxides,\cite{Wu/Khomskii:2007,Maitra/Valenti:2007} 
we calculate the DOS with spin-orbit interactions (SOI) [\autoref{fig:dos}(b)].
Although the SOI enable the spin-up and spin-down bands to mix, we 
find an electronic structure similar to that without the interactions. 
The Ru 4$d$-band occupation at $E_\mathrm{F}$ is weakly modified and the gap in the 
conduction band is slightly reduced.
Nonetheless, the weakly dispersed $t_{2g}$ states persists, and {\scro}
remains metallic [$N(E_\mathrm{F}$)$\sim4.36$~eV/Ru atom] 
consistent with SOI calculations on {bulk} SrRuO$_3$,\cite{Rondinelli/Spaldin_et_al:2008} where 
spin-orbit coupling also does not greatly alter the electronic structure.\\

\noindent\textbf{Origin of acentricity.}
For purposes of evaluating the design approach and the proposed geometric-induced 
inversion-symmetry-breaking displacements, we perform a group theoretical analysis\cite{Aroyo/Perez-Mato_et_al:2006,Orobengoa/Perez-Mato:2009}  
of the $Pmc2_1$ structure with respect to a fictitious 
prototypical centrosymmetric phase ($P4/mmm$ symmetry), 
reducing the polar structure
into a set of symmetry-adapted modes  
associated with  different irreducible representations (irreps) 
of the $P4/mmm$ phase 
(see Supplementary Figure 1 and Supplementary Table 4). 
Note that the hypothetical structure would be  
related to an experimentally accessible high-temperature phase, 
assuming that the crystal would not first decompose. 
The normalized distortion vector is
 $\xi \simeq 0.32\Gamma_5^- + 0.54M_2^+ + 0.78M_5^-$, where 
$\Gamma_5^-$ corresponds to a polar $ir$-active mode, and $M_2^+$ 
and $M_5^-$ describe $k\neq0$ in-phase and out-of-phase rotation modes, respectively.

\begin{figure}
\centering
\includegraphics[width=0.98\columnwidth,clip]{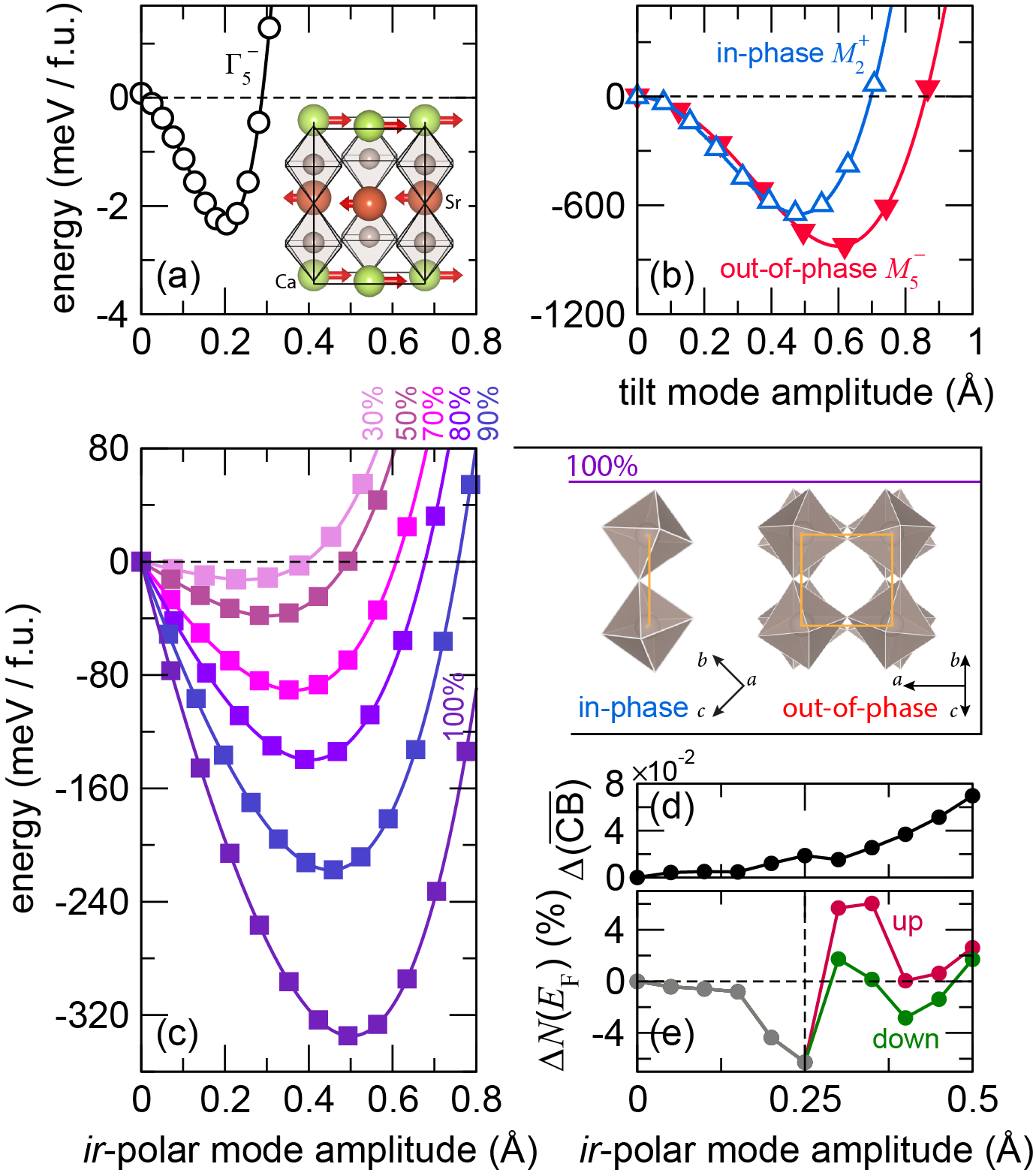}\vspace{-0.5\baselineskip}
\caption{%
\textbf{Electron--lattice interactions on the ground state stability.}
(a) Polar ($ir$-active) mode with $\Gamma_5^-$ symmetry.
The inset depicts the polar cation displacement pattern; 
apical oxygen and (small) Ru displacements are omitted for clarity.
(b)
In-phase and out-of-phase RuO$_6$ rotational modes. 
(c) 
Normalized energy gain obtained by increasing amplitude of the polar 
displacements at fixed percentages of the total $a^+b^-b^-$ tilt pattern 
found in the ground-state structure, {i.e.}, described by the combination $M_2^+ \oplus M_5^-$.
The inset depicts the 100\% rotational distortions.
(d) Change in the position of the Sr and Ca conduction band center of mass ($\Delta\overline{\mathrm{CB}}$) and (e) variation in the 
number of states at the Fermi level [$\Delta N(E_\mathrm{F})$] 
normalized to the value without the polar distortion.
The ferromagnetic state is stable for polar displacements larger than $\sim$0.25~\AA.
}
\label{fig:polar_wrt_rots}
\end{figure}

The amplitude and energetic contribution of each irrep in the polar 
structure provides insight into the stability of the NSCM (\autoref{fig:polar_wrt_rots}).
Although the zone-center polar mode, transforming as irrep $\Gamma_5^-$ 
is  weakly unstable 
in the high-symmetry $P4/mmm$ structure (see Supplementary Figure 2 and  Supplementary Table 5), {i.e.},  
characterized by a negative quadratic energy surface with respect to the 
increasing amplitude of the polar distortion [Fig.~\ref{fig:polar_wrt_rots}(a)], 
its energetic contribution to the ground state structure is two orders of magnitude 
smaller than that obtained from the octahedral rotation modes 
[\autoref{fig:polar_wrt_rots}(b)].
To show that the geometric RuO$_6$ rotations drive the polar Sr and Ca displacements 
[\autoref{fig:polar_wrt_rots}(a,inset)], circumventing the contraindication 
between metallicity and the $ir$-mode,  
we map the total energy evolution of {\scro} with increasing amplitude 
of the $ir$-polar  mode at different fixed values of the $a^+b^-b^-$ tilt pattern 
described by ($M_2^+ \oplus M_5^-$) [\autoref{fig:polar_wrt_rots}(c)].
We find that with increasing amplitude of the 
$a^+b^-b^-$ tilt pattern, the negative curvature about the origin vanishes;  
the energy landscape evolves to a parabola with a single energy minimum, indicating 
the hardening (stabilization) of the polar phonon mode at finite amplitude 
induced by the geometric RuO$_6$ octahedral rotations, which 
fully accounting for the loss of inversion symmetry.

\autoref{fig:polar_wrt_rots}(d)  shows there are minor changes in the position of the 
band center of mass for the principal atoms 
involved in the polar mode, {i.e.}, the Ca ($4s3d$)  and Sr ($5s4d$) states 
which are located $\sim$6.2~eV above $E_\mathrm{F}$ in the conduction band.
The relative change in the number of states at the Fermi level 
$[\Delta N(E_\mathrm{F})$, \autoref{fig:polar_wrt_rots}(e)] is also small 
with respect to the centrosymmetric structure  and  increasing amplitude of the polar 
mode, confirming the weak interaction between the inversion 
symmetry lifting displacements and the low-energy electronic structure.
The conduction bands marginally shift to higher energy ($\sim$0.06~eV) 
and the change in $N(E_\mathrm{F})$ is approximately $\pm$5\%, validating the 
proposed NCSM design guidelines.
Interestingly, the polar mode activates the ferromagnetic ground state---observed by 
the exchange split electronic structure that appears around $\sim$0.25~{\AA} 
in  \autoref{fig:polar_wrt_rots}(e), indicative of large spin-phonon coupling\cite{Lee/Rabe:2010,Zayak/Rabe:2008}. \\

\begin{figure}
\centering
\hspace*{-9pt}  
\includegraphics[width=0.50\textwidth,clip]{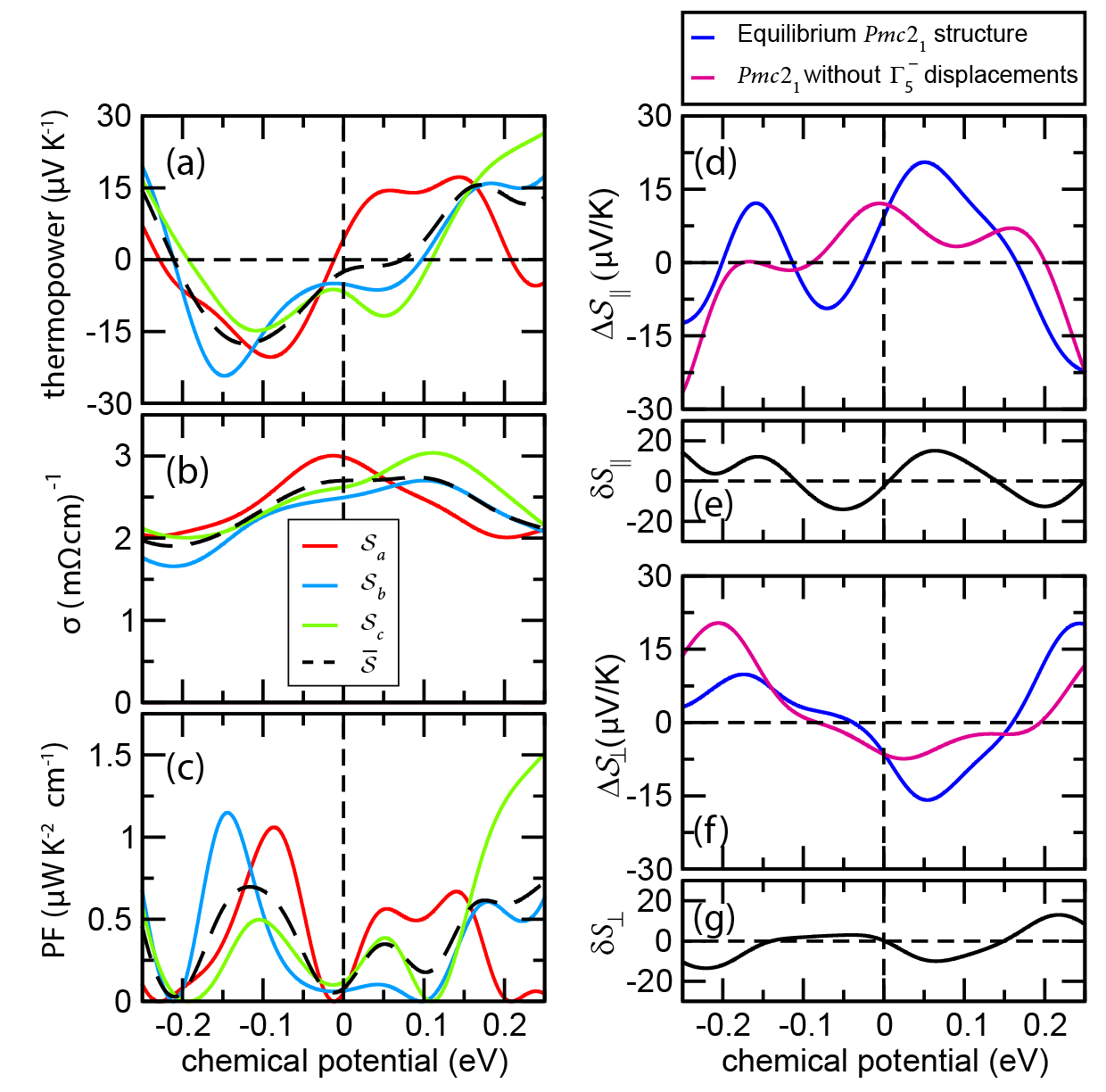}\vspace{-0.9\baselineskip}
  \caption{\textbf{Chemical potential-dependent properties and anisotropy at 300~K.}
\scro transport properties (left panels) 
   as a function of the chemical potential, $\mu$: 
  (a) thermopower $\mathcal{S}$, (b) conductivity $\sigma$, 
  and (c) the power factor $\mathrm{PF}$.
  Red, blue, and green line are the transport properties along the $a$, $b$,  
  and the polar $c$-axis, respectively. The dashed line indicates the total  	  behavior, given as an average of the three principle components.
   Anisotropy (right panels) 
  along (d) the $ab$-plane ($\Delta\mathcal{S}_\parallel$) with (blue) and without (magenta) polar displacements;  the difference of the two curves is depicted in (e).  
  (f) The anisotropy between the polar $c$-axis and the $ab$-plane  ($\Delta\mathcal{S}_\perp$), and (g) 
  the difference between the two curves. 
\label{fig:300K}}
\end{figure}

\noindent\textbf{Thermoelectric response.}
Owing to the orthorhombicity   in \scro, the recent interest in 
thermoelectric properties of perovskite oxide superlattices,\cite{garcia:2012} 
and the fact that degenerately doped ferroelectrics with polar axes exhibit 
unusual thermoelectric properties,\cite{Lee:2012,Choy:1992,Kolodiazhnyi:2003} 
we evaluate the thermopower anisotropy at 300~K using Botlzmann transport theory within 
the constant scattering time approximation.

\autoref{fig:300K} shows the relevant thermoelectric properties in 
the chemical potential range  varying from -0.25~eV to 0.25~eV.
The average  thermopower $\bar{\mathcal{S}}$, denoted by the broken line in \autoref{fig:300K}(a),  
ranges from 14 to -18$~\mu \textrm{V K}^{-1}$ with 
two sign changes near $\mu\simeq-0.21$ and $\mu\simeq0.07$~eV, 
indicating a change in the dominant charge carriers from holes to electrons, and then 
from electrons to holes, respectively. 
We also note that above $\mu\simeq0.16$~eV,  the thermopower components along the polar $c$ direction dominate the thermal properties.
The conductivity $\sigma$  is found to oscillate in a narrow window over the 
chemical potential range considered, 
and all the three components resolved along the crystallographic exhibit 
similar behavior [\autoref{fig:300K}(b)]. 
Above $\mu\!\simeq\!0.05$~eV, 
where the Seebeck coefficient along the polar axis is extremized, 
the conductivity along the $c$-axis ($\sigma_c$) begins to exceed that along the  
other directions.

\autoref{fig:300K}(c) shows the total and crystal axis-decomposed power factor 
$\mathrm{PF}\!\!=\!\!\sigma\mathcal{S}^2$.
The total power factor exhibits a pronounced local
maximum, $\mathrm{PF}\!\!\simeq\!\!0.7~\mu\textrm{W K}^{-2}\!\textrm{ cm}^{-1}$, near $\mu$=-0.12~eV  (hole doping $h$=3$\times$10$^{21}~$cm$^{-3}$), and 
two other less pronounced local maxima, $\mathrm{PF}\!\simeq\!0.35~\mu\textrm{W K}^{-2}\!\textrm{ cm}^{-1}$ 
at $\mu\!\sim\!0.05$~eV and $\mathrm{PF}\!\simeq\!0.61~\mu\textrm{W K}^{-2}\!\textrm{ cm}^{-1}$ 
at $\mu\!\sim\!0.18$~eV, 
which are negligible when compared to thermoelectric materials found in 
applications (20-50~$\mu\textrm{W K}^{-2}\!\textrm{ cm}^{-1}$).\cite{Synder/Toberer:2008}
It is interesting, however, that the maximum at $\mu\!\sim\!0.18$~eV (electron doping of $n=6\times10^{21}\textrm{ cm}^{-3}$)
is in the region where the Seebeck coefficient and the conductivity along the polar axis both dominate the transport behavior along the other directions.
To show that this anisotropy arises from the polar displacements and 
not  the orthorhombic lattice symmetry of (Sr,Ca)Ru$_2$O$_6$,
we calculate the thermopower anisotropy in the $ab$-plane as  
$\Delta\mathcal{S}_{\parallel}=\mathcal{S}_{a}-\mathcal{S}_{b}$ [\autoref{fig:300K}(d)], 
and the thermopower anisotropy along the {polar} $c$-axis as 
$\Delta\mathcal{S}_\perp=\mathcal{S}_{c}-\frac{1}{2}(\mathcal{S}_{a}+\mathcal{S}_{b}$) [\autoref{fig:300K}(f)], 
for the equilibrium $Pmc2_1$ structure and for a hypothetical structure with the 
polar atomic displacements described by irrep $\Gamma_{5}^{-}$ fully removed.
When we remove the polar displacements, we find that 
$\Delta\mathcal{S}_{\parallel}$  is modified for all chemical potentials examined except 
at 0~eV.
%
%
The difference between the two cases $\delta S _\parallel$  
ranges between $\pm$15~$\mu \textrm{V K}^{-1}$  [\autoref{fig:300K}(e)].   
The thermopower anisotropy along the polar $c$-axis, $\Delta\mathcal{S}_\perp$, 
behaves as $\Delta\mathcal{S}_{\parallel}$, except for a sign change at 0~eV: 
 %
%
%
The  difference in the $c$-axis anisotropy, $\delta S _\perp$, 
also ranges between $\pm$15~$\mu \textrm{V K}^{-1}$  for the two structures [\autoref{fig:300K}(g)].
Consistent with our design guidelines, at $\mu=0$~eV, the change in the 
thermopower anisotropy is essentially zero, indicating weak coupling between 
the states at the Fermi level and the polar atomic displacements that 
lift inversion symmetry.
The role of the polar distortions becomes even more evident by analyzing the anisotropy change in the $bc$-plane [see Supplementary Figure 3] where the lattice 
parameters are identical.  The removal of polar displacements is found to produce a shift relative to the chemical potential below $\mu\!=\!0.18$~eV. Above this value, we find that the  anisotropy in the $bc$-plane is small in the structure without the polar distortions, whereas it is maximized in the equilibrium structure. Such behavior is also observed in \autoref{fig:300K}(f) over the same electron-doped region. The magnitude of the anisotropy, $|\delta S _\perp|$, in the ground state structure exceeds that of the phase with the polar displacements removed.
%

%
\begin{figure}
\centering
\includegraphics[width=0.99\columnwidth,clip]{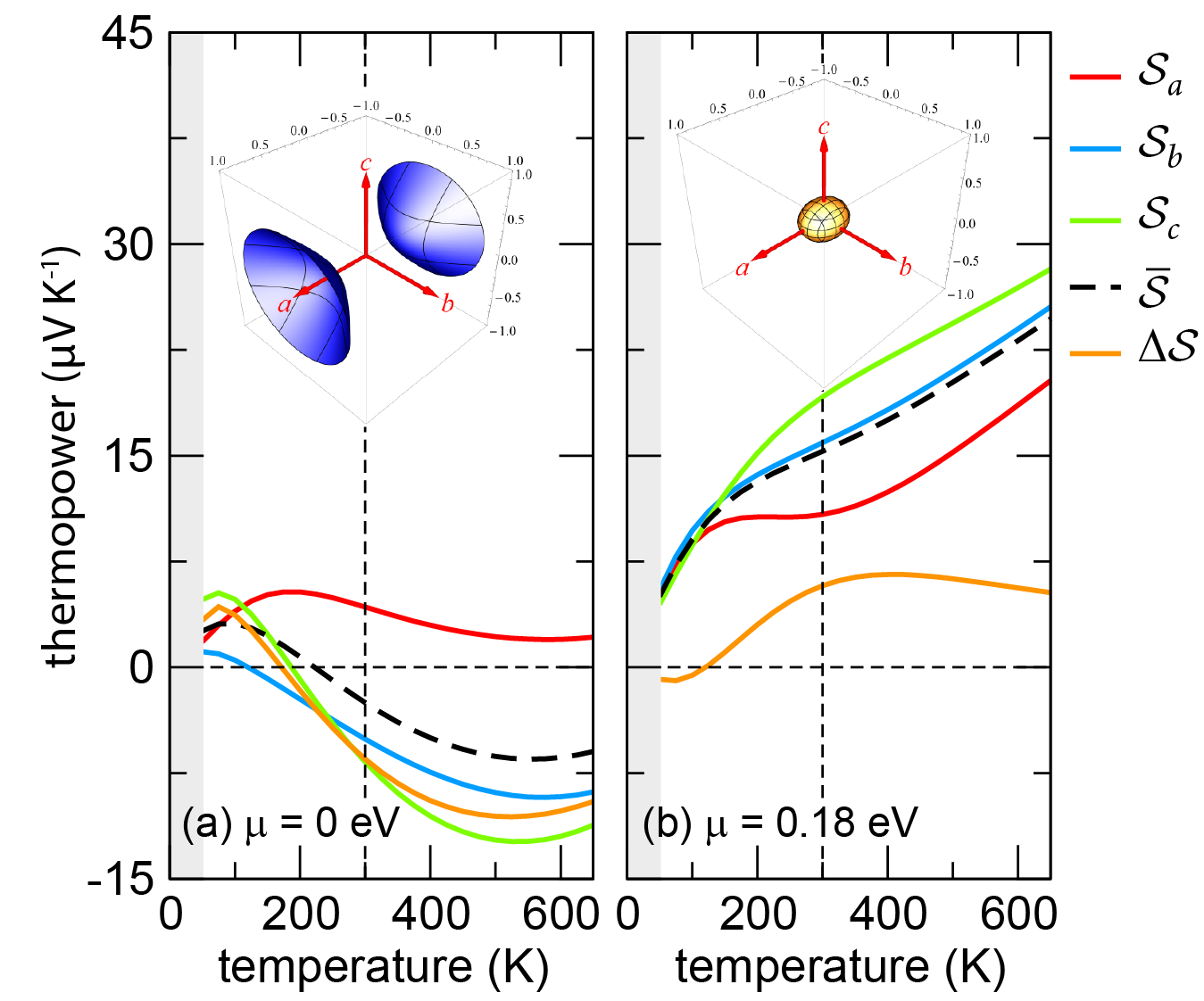}\vspace{-0.8\baselineskip}
\caption{\textbf{Temperature-dependent thermopower in paramagnetic \scro.} Temperature-dependent thermopower along the 
$a$, $b$ and $c$ axes at (a) $\mu$=0~eV and (b) $\mu$=0.18~eV. 
The shaded area indicates the ferromagnetic phase below the calculated Curie temperature $T_\mathrm{c}\!\sim\!50$~K.
The average thermopower is calculated as
$\bar{\mathcal{S}}=(\mathcal{S}_a + \mathcal{S}_b +\mathcal{S}_c)/3$.
The insets depict the 300~K quadric surface representations of the second-rank 
Seebeck tensor, revealing enhanced anisotropy upon doping along the polar axis.}
\label{fig:anisotropy_vs_t}
\end{figure}

\autoref{fig:anisotropy_vs_t} shows the computed temperature-dependent 
thermopower $\mathcal{S}$ as a function of chemical potential $\mu=0$ and 0.18~eV,  
which is selected because the power factor ($\sigma\mathcal{S}^2$) exhibits a 
 local maximum that is due mainly to the Seebeck coefficient 
along the polar axis, $\mathcal{S}_c$ [\autoref{fig:300K}(c)].
Due to identical lattice constants along the $b$ and $c$ axes in \scro, 
 $\mathcal{S}_b$ and $\mathcal{S}_c$ have similar behavior for $\mu=0$~eV. 
The total thermopower, $\bar{\mathcal{S}}$, ranges from $\sim\!2.5$ to -7$\!~\mu \textrm{V K}^{-1}$, and exhibits parabolic behavior  with a maximum around 550~K.
%
The thermopower anisotropy along the {polar} $c$-axis, 
$\Delta\mathcal{S}_\perp$ exhibits the same behavior. At $\mu=0$~eV, the 
thermopower anisotropy is relatively large 
($\left|\Delta\mathcal{S}_\perp\right|\!\sim\!6.3\!~\mu \textrm{V K}^{-1}$), reflected in 
the 300~K representation of the second-rank Seebeck tensor (inset) appearing 
as a hyperboloid with two sheets.
%

For $\mu\!=\!0.18$~eV the scenario is  different. 
$\bar{\mathcal{S}}$ has a  quasi-linear behavior from low to high temperature: 
$\mathcal{S}_a$, $\mathcal{S}_b$ and $\mathcal{S}_c$ all become positive.
Over the entire temperature range considered, the thermopower along the polar axis  $\mathcal{S}_c$ dominates. 
At 300~K (broken vertical line)  the second-rank Seebeck tensor quadric surface becomes an elongated ellipsoid, flattened along the $c$ direction.
Accordingly the thermopower anisotropy at 300~K becomes $\sim\!5.8\!~\mu \textrm{V K}^{-1}$.\\

\begin{figure}
\centering
\hspace*{-3pt}\includegraphics[width=0.49\textwidth,clip]{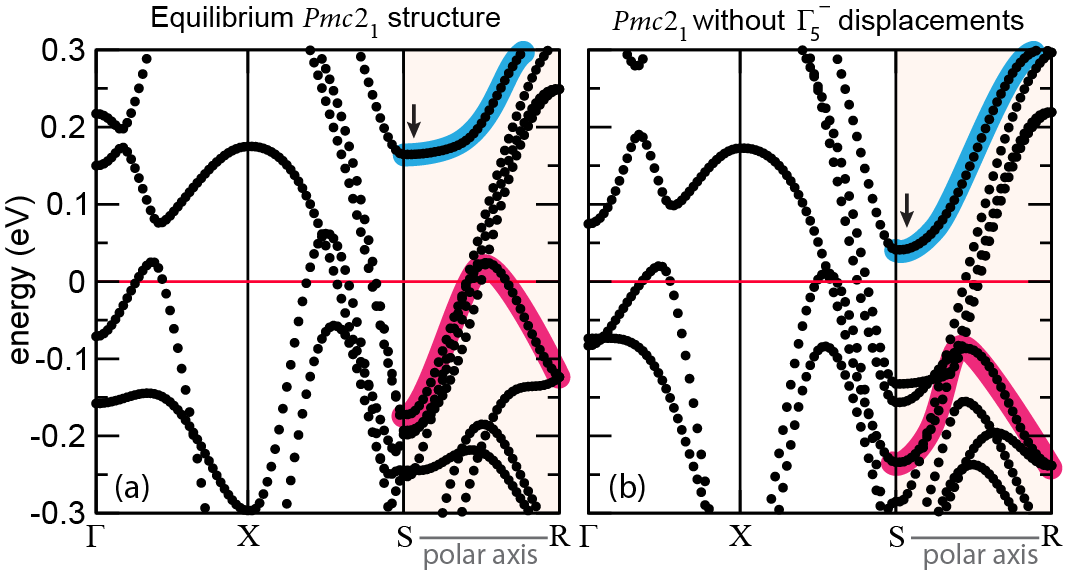}\vspace{-0.8\baselineskip}
  \caption{\textbf{Effect of  polar displacements on the electronic structure}. 
In the left (right) panel the bands structure for the non-magnetic layered
(Sr,Ca)Ru$_{2}$O$_{6}$ (without $\Gamma_{5}^{-}$ displacements). 
The highlighted bands are mainly affected by the polar displacements. 
The red (lower energy) bands are 
partially occupied in the equilibrium structure. The shaded path 
corresponds to a $k$-space trajectory parallel to the polar axis.
Brillouin zone labels: 
$\Gamma$ (0, 0, 0), X ($\pi$/2, 0, 0), S ($\pi$/2, $\pi$/2, 0), R ($\pi$/2, $\pi$/2, $\pi$/2).}
  \label{fig:bands}
\end{figure}  

\noindent{\sffamily \bfseries Discussion}\\
The thermoelectric behavior as a function of doping 
can be understood by examining the electronic band structures 
of the equilibrium $Pmc2_1$ structure with that of the hypothetical (Sr,Ca)Ru$_2$O$_6$ 
system with the polar atomic displacements removed (\autoref{fig:bands}). 
Inspection of the electronic band structure along the lines of 
symmetry $X\!-\!S$ (corresponding to $b$ in real space) and $S\!-\!R$ (corresponding to the polar $c$ axis)
reveal strongly hybridized Ru $4d$ -- O $2p$ bands dispersing from -0.2~eV to 0.3~eV, 
where the electron contribution dominates until $\sim$0.07~eV above  the Fermi level [\autoref{fig:bands}(a)], 
explaining the sign change of  $\bar{\mathcal{S}}$ near $\mu\simeq0.07$~eV in \autoref{fig:300K}(a).

Additionally, the polar displacements largely modify the band structure along $S\!-\!R$ 
in the energy range 0.05-0.30~eV. 
In the absence of the polar mode, 
the band near 0.16~eV (arrow in \autoref{fig:bands}), 
which is derived from the oxygen $2p$ states in the CaO and SrO planes, shifts to lower 
energy and broadens. 
In the equilibrium structure, this band along $S\!-\!R$ is higher in energy than 
the band which changes dispersion along  $\Gamma\!-\!X$ (energy range 0.05-0.18~eV); 
however, when the polar displacements are removed, the oxygen-derived 
band along $S-R$ becomes lower than that of the band center  of mass along $\Gamma\!-\! X$.
It is the dependence of these two bands on the amplitude of the polar atomic 
displacements that leads to the different behavior appearing in \autoref{fig:300K}(f)  
for $\mu=0.18$~eV.
We find that the 300~K thermopower anisotropy of \scro at $\mu$=0.18~eV is close to  that of {centrosymmetric} 
YBa$_2$Cu$_3$O$_{7-\delta}$ (YBCO, $\sim\!7$-$10~\mu \textrm{V K}^{-1}$), \cite{kaiser:1991,zeuner:1995} which exhibits strong thermal and electrical anisotropy owing to the two-dimensional (2D) crystal structure.
The thermopower anisotropy in other metallic systems, 
{e.g.}, the 
hexagonal delafossite oxides PdCoO$_2$ and PtCoO$_2$,\cite{ong:2010} 
also derive from the 2D layered topology of the system, and so 
present large Seebeck coefficients along the trigonal axis, {i.e.}, 
the axis which is most insulating.
In contrast, the thermopower anisotropy in (Sr,Ca)Ru$_2$O$_6$ 
is not related to the dimensionality of the system, but rather to the existence of a polar axis 
along which the electrical conductivity is {highest}.
The peculiar thermopower anisotropy in  (Sr,Ca)Ru$_2$O$_6$ 
dictates that the electric field resulting from an applied heat flux 
to the material will be non-collinear. 
This property is a fundamental feature for any anisotropic thermoelectric devices. \cite{burkov:1995} 
It enables the heat flux to be measured in a geometry {perpendicular}  
to the induced electrical current, specifically at locations 
where the temperatures are equal (see Supplementary Figure 4 and Supplementary Note 1). 
%
%
Given the metallic conductivity of \scro and that the relaxation time is 
small in comparison with that of semiconductors or 
insulators already finding use  in thermoelectric devices, this polar-NCSM
would enable sensing on sub-nanosecond timescales (comparable to  YBCO\cite{zeuner:1995}).
New applications of these oxide materials could be found in 
ultrafast-thermoelectric devices,\cite{burkov:1995,zeuner:1995} where 
stability under extreme conditions, speed, and the ability to 
measure  heat  fluxes of high density are key 
requirements, {e.g.}, thermal (heat) radiation detectors.
Moreover, because of the compatibility of perovskite oxides 
with Si-based CMOS technologies,\cite{Warusawithana/Schlom:2009} 
we anticipate these designed NCS metals will more readily find integration and device 
development than previously identified materials with large anisotropic thermoelectric 
responses. 

In summary, we designed a polar-noncentrosymmetric
ruthenate conductor and described how to eliminate
the incompatibility between metallicity and acentricity
in complex transition metal oxides. 
We articulated a
paradigm for which new noncentrosymmetric metals may
be found, and showed that (Sr,Ca)Ru$_2$O$_6$ exhibits
Seebeck coefficients with large anisotropy derived from
the polar structure. 
We believe that other
polar metals could exhibit such anisotropic thermal properties; the important
discriminating feature, however, is the $k$-space evolution of the electronic structure 
with the real space polar distortion. 
Broadly, such thermopower anisotropy may also be 
found in any NCS metal, e.g., in compounds which are chiral but non-polar, 
but continued studies are required to clarify the extent of the response. 
We hope this work motivates the
synthesis of new materials, and the discussion of new
applications where highly anisotropic thermoelectric
responses in metals can be leveraged.\\

\noindent{\sffamily \bfseries Methods}\\
\noindent{\bfseries \emph{Ab initio} calculations and group theoretical analysis}\\
We perform first-principles density functional calculations 
within the local spin density approximation (LDA),  which is known to provide 
a reliable description of perovskite ruthenates\cite{Zayak/Rabe:2006,Zayak/Rabe:2008,Rondinelli/Spaldin_et_al:2008} 
as implemented in the Vienna
{\it Ab initio} Simulation Package ({\sc vasp}). 
\cite{Kresse/Furthmuller:1996a,Kresse/Joubert:1999} 
Calculations carried out with the Perdew-Burke-Ernzerhof 
generalized gradient approximation functional revised for solids (PBEsol) \cite{PBEsol:2008} 
provide essentially the same results  (see Supplementary Table 6).
We use the projector augmented wave (PAW) method \cite{Blochl:1994} 
to treat the core and valence electrons using a 550~eV plane wave expansion 
and the following  electronic configurations:  
$4s^24p^65s^2$ (Sr), 
$3p^64s^2$ (Ca), 
$4p^65s^24d^6$ (Ru), 
and $2s^22p^4$ (O). 
A $7\times7\times5$ 
Monkhorst-Pack $k$-point mesh\cite{Monkhorst/Pack:1976} 
and Gaussian smearing (20~meV width) was used 
for the Brillouin zone (BZ) sampling and integrations. 

For structure optimization, 
we relax the atomic positions to have forces less than 0.1
meV~\AA$^{-1}$. 
We impose as a constraint $b=c$ 
(note that in bulk ruthenates the deviations between the ``short'' axes  are small\cite{Zayak/Rabe:2008}) 
to simulate the situation under thin film growth on a cubic substrate with a square lattice net. 
We search for the ground state structure by computing the 
phonon frequencies  and eigenmodes of the high-symmetry 
centrosymmetric (Sr,Ca)Ru$_2$O$_6$ structure (space group 
$P4/mmm$), and then systematically ``freeze-in'' linear 
combinations of the unstable modes, performing 
structural optimization on these candidate low-symmetry structures 
until the global minimum is obtained.
The group theoretical analysis was aided with the
\textsc{isodistort}\cite{Campbell:wf5017} and \textsc{amplimodes}\cite{Orobengoa:ks5225,Perez-Mato:sh5107} software.\\

\noindent{\bfseries Thermoelectric properties}\\
For the calculation of the Seebeck coefficient and other transport properties, we use Boltzmann transport
theory\cite{Ashcroft/Mermin:Book, Ziman:Book} within the 
constant scattering time ($\tau$) approximation as implemented in the 
\textsc{boltztrap} code.\cite{Madsen:2006}
We perform a non-self-consistent band-structure calculation with a much denser 
sampling of 27$\times$33$\times$33 (4046 $k$-points in the irreducible Brillouin zone; 
29403 $k$-points in the full Brillouin zone), using the equilibrium ground state structure. 
The Seebeck coefficients $\mathcal{S}$ are calculated independently from $\tau$.
The electrical conductivity, however, requires a known value of $\tau$, and so 
we introduce an empirical parameter.
Here, the relaxation time value $\tau = 0.23\times10^{-14}$~s is 
employed in all calculations, which we obtain from fitting the
room-temperature conductivity\cite{Keawprak:2012} $\sigma\sim$2.7 $\times$ 10$^5$~S~\!m$^{-1}$ of 
solid solution Sr$_{0.5}$Ca$_{0.5}$RuO$_{3}$.
We believe this estimate to provide a meaningful approximation based on our 
excellent agreement with other physical properties of the solid solution reported in 
the main text.
The thermoelectric properties at room temperature are calculated using the 
zero Kelvin polar structure based on the fact that the high-temperature structural transitions 
away from orthorhombic symmetry in bulk CaRuO$_3$ and SrRuO$_3$ 
do not occur until 1550~K\cite{Ranjan:2007} and 685~K, respectively.\cite{Brendan:2002}
Based on our calculated magnetic Curie temperatures of $\sim$50~K, 
we also compute all thermoelectric responses using a non-magnetic configuration.\\
%

\noindent{\sffamily \bfseries Acknowledgements}\\
This work was supported by the Army Research Office under grant no.\ 
W911NF-12-1-0133.
The authors acknowledge the High Performance Computing Modernization Program 
(HPCMP) of the DOD for providing computational resources that have contributed to 
the research results reported herein.
The authors thank K.R.\ Poepplemeier and C.J.\ Fennie 
for insightful discussions.\\

\noindent{\sffamily \bfseries Author contributions}\\
The study was planned, calculations carried out, and the manuscript 
prepared by D.P. and J.M.R.
Both authors discussed the results, wrote, and commented on the 
manuscript.\\

\noindent{\sffamily \bfseries Additional information}\\
Supplementary information is available in the online version of the paper. 
Reprints and permissions information is available online at
\url{www.nature.com/reprints}. 
Correspondence and requests for materials should be addressed to 
J.M.R.\\

\noindent\textbf{Competing financial interests:}
The authors declare no competing financial interests.

%


\end{document}